\documentclass[preprint]{aastex}
\usepackage{float,graphicx,amsmath,multirow}
\usepackage{color}

\title{CSO and CARMA Observations of L1157.  I.  A Deep Search for Hydroxylamine (NH$_2$OH)}
\author{Brett A. McGuire}
\affil{National Radio Astronomy Observatory, Charlottesville, VA 22903\\ Division of Chemistry and Chemical Engineering, California Institute of Technology, Pasadena, CA 91125}
\author{P. Brandon Carroll}
\affil{Division of Chemistry and Chemical Engineering, California Institute of Technology, Pasadena, CA 91125}

\author{Niklaus M. Dollhopf}
\affil{Department of Astronomy, University of Virginia, Charlottesville, VA 22903 \\National Radio Astronomy Observatory, Charlottesville, VA 22903}

\author{Nathan R. Crockett}
\affil{Division of Geological and Planetary Sciences, California Institute of Technology, Pasadena, CA 91125}

\author{Joanna F. Corby}
\affil{Department of Astronomy, University of Virginia, Charlottesville, VA 22903 \\National Radio Astronomy Observatory, Charlottesville, VA 22903}

\author{Ryan A. Loomis}
\affil{Department of Astronomy, Harvard University, Cambridge, MA 02138}

\author{Andrew Burkhardt}
\affil{Department of Astronomy, University of Virginia, Charlottesville, VA 22903 \\National Radio Astronomy Observatory, Charlottesville, VA 22903}

\author{Christopher Shingledecker}
\affil{Department of Chemistry, University of Virginia, Charlottesville, VA 22903}

\author{Geoffrey A. Blake}
\affil{Division of Chemistry and Chemical Engineering, California Institute of Technology, Pasadena, CA 91125 \\Division of Geological and Planetary Sciences, California Institute of Technology, Pasadena, CA 91125}
\and
\author{Anthony J. Remijan}
\affil{National Radio Astronomy Observatory, Charlottesville, VA 22903}

\begin{document}

\begin{abstract}

A deep search for the potential glycine precursor hydroxylamine (NH$_2$OH) using the Caltech Submillimeter Observatory (CSO) at $\lambda = 1.3$ mm and the Combined Array for Research in Millimeter-wave Astronomy (CARMA) at $\lambda = 3$ mm is presented toward the molecular outflow L1157, targeting the B1 and B2 shocked regions.  We report non-detections of NH$_2$OH in both sources.  We a perform non-LTE analysis of CH$_3$OH observed in our CSO spectra to derive kinetic temperatures and densities in the shocked regions.  Using these parameters, we derive upper limit column densities of NH$_2$OH of $\leq1.4 \times 10^{13}$~cm$^{-2}$ and $\leq1.5 \times 10^{13}$~cm$^{-2}$ toward the B1 and B2 shocks, respectively, and upper limit relative abundances of $N_{NH_2OH}/N_{H_2} \leq1.4 \times 10^{-8}$ and $\leq1.5 \times 10^{-8}$, respectively.

\end{abstract}

\section{Introduction}

Glycine (NH$_2$CH$_2$COOH), the simplest amino acid, has been identified in cometary \citep{Elsila2009} and meteoritic samples \citep{Kvenvolden1970}, however it is not yet known whether the species is formed in these solid bodies, in the interstellar medium (ISM), or both. Indeed, a detection in the ISM has so far proven ellusive \citep{Snyder2005,Cunningham2007,Jones2007}.  Although chemical models predict the formation of glycine in the ISM, in low abundance (see, e.g. \citet{Garrod2013} and refs. therein), it has been difficult to observationally constrain them without a detection of the species.  The relative likelihood of different possible formation routes can, however, be constrained by careful observation of the reactants used in the chemical models to form glycine.

One pathway which has garnered significant interest is the formation of glycine through the reaction of hydroxylamine (NH$_2$OH), or its protonated and ionized derivatives, with acetic acid (CH$_3$COOH), a known interstellar molecule \citep{Mehringer1997}.  Ionization or protonation of NH$_2$OH under interstellar conditions should be efficient \citep{Angelelli1995,Boulet1999}, and subsequent laboratory work has demonstrated the formation of glycine from NH$_2$OH and its ionized and protonated forms \citep{Blagojevic2003,Snow2007}.  Although recent theoretical work has suggested that these gas-phase routes forming glycine through these reactions are inefficient under interstellar conditions \citep{Barrientos2012}, condensed-phase surface-mediated reactions remain a possibility, and observational constraints of the precursors are still desirable for the refinement of models.  While acetic acid is readily-constrained, NH$_2$OH has yet to be observed in the ISM.

In recent years, laboratory work has shown that a number of pathways exist which result in efficient formation of NH$_2$OH on grain surfaces.  \citet{Zheng2010} demonstrated the formation of NH$_2$OH in H$_2$O-NH$_3$ ices after UV-irradiation.  Formation by successive hydrogenation of NO has been shown to proceed efficiently, barrierlessly, and in high yield \citep{Congiu2012a,Congiu2012b,Fedoseev2012,Ioppolo2014,Minissale2014}.  Most recently, \citet{He2015} demonstrated the efficient formation of NH$_2$OH via oxidation of NH$_3$ in ices with a low barrier to activation (of order 1000 K).

Earlier modeling work by \citet{Garrod2008} also suggested that NH$_2$OH is formed in high abundance in interstellar ices, and is subsequently liberated into the gas-phase during the warm-up period of emerging hot cores and hot corinos in readily-detectable quantities.  Observational efforts by \citet{Pulliam2012}, however, failed to detect NH$_2$OH toward a selection of such sources known to be rich in complex organic material (e.g. Sgr B2(N) and Orion-KL) establishing upper limits of N$_{NH_2OH}/N_{H_2} < 10^{-9} - 10^{-11}$.  Refinement of the \citet{Garrod2008} model in subsequent work \citep{Garrod2013}, as well as limiting modeling presented in the aforementioned laboratory studies, resulted in predicted gas-phase abundances in line with the upper limits established by \citet{Pulliam2012}.  Yet, with the inclusion of the H + HNO $\rightarrow$ HNOH pathway into the model, as described by the laboratory work of \citet{Congiu2012a}, condensed phase abundances of NH$_2$OH again approach N$_{NH_2OH}/N_{H_2} \sim 10^{-6}$, with gas-phase abundances of N$_{NH_2OH}/N_{H_2} \sim 10^{-7}$.

The \citet{Garrod2013} model largely assumes that the release of NH$_2$OH into the gas-phase is a gradual process dominated by the warm-up of the hot core.  Both the laboratory work and the models, however, predict NH$_2$OH is initially formed in large abundance at very cold ($<20$ K) temperatures and early in the evolution of these sources.  Thus, the most optimistic source for a detection of gas-phase NH$_2$OH is one where the reservoir of condensed-phase NH$_2$OH formed at low temperatures is liberated \textit{en masse} into the gas-phase prior to release by thermal mechanisms.

Shocked regions displaying high degrees of molecular complexity likely represent this best-case scenario.  In these regions, complex molecules are formed efficiently in ices at low temperatures, but are not otherwise liberated into the gas-phase except by thermal desorption at much greater temperatures.  When these ices are subjected to shocks, however, the mantle is non-thermally ejected into the gas-phase, resulting in large abundances of relatively cool ($T_{rot}<100$ K), complex organic material (see, e.g., \citet{Torres2006}).  One of the most prominent of these regions is the young protostellar outflow L1157. Numerous recent studies report high degrees of molecular complexity arising from shocked regions within the outflow, which originates in cold, quiescent gas around the protostar (see, e.g., \citet{Arce2008} and \citet{Codella2015}).

Here, we present deep searches for NH$_2$OH using the Caltech Submillimeter Observatory (CSO) at $\lambda=1.3$ mm and the Combined Array for Research in Millimeter-wave Astronomy (CARMA) at $\lambda=3$ mm toward the L1157 outflow.  We report non-detections of NH$_2$OH in both searches.  In order to derive upper limits to the column density of NH$_2$OH, we use transitions of CH$_3$OH observed with the CSO to constrain the kinetic temperature and density in the shocked gas using a radiative-transfer approach, and CARMA images of CH$_3$OH to determine the size of the shocked gas.  Finally, we estimate upper limits to the abundance of NH$_2$OH and discuss possible implications.

\section{L1157}

L1157 is a dark cloud in Cepheus located a distance of $\sim$250 pc \citep{Looney2007}, and contains a prototypical shocked bipolar outflow from a Class 0 protostar.  It has been the subject of great interest in the last twenty years, with numerous studies investigating the physical conditions within the source.  Due to the variety of methods used for these studies, direct comparisons between results is challenging.  Nevertheless, an overall picture does come into view, and this general description will be sufficient for the discussion presented here.   

Originating in the cold, quiescent gas ($T \sim 13$ K, \citet{Bachiller1993}) surrounding the Class 0 protostar L1157-mm, the southern lobe of the accelerated outflow  ($T \sim 50 - 100$ K, \citet{Bachiller1993}) has undergone two major shocks, referred to as L1157-B1 and L1157-B2.  The B1 shock is younger and warmer ($\sim$2000 yr, $T_{kin} \sim 80 - 100$ K) than the B2 shock ($\sim$4000 yr, $T_{kin} \sim 20 - 60$ K), and many complex chemical species are observed in enhanced abundance toward both shocks due to non-thermal desorption from grains \citep{Mendoza2014,Codella2015}.  While the absolute values of these physical parameters vary somewhat within the literature, this qualitative picture and enhancement in chemical abundance is consistently reported.

\section{CSO Observations}

The spectrum toward L1157 obtained with the CSO was collected over 8 nights in 2012 July, August, and September, and 7 nights in 2014 September and December.  The telescope was pointed at the B1 shocked region at $\alpha$(J2000)~=~20$^{\mbox{h}}$39$^{\mbox{m}}$07$^{\mbox{s}}$.7, $\delta$(J2000)~=~68$^{\circ}$01$^{\prime}$15$^{\prime\prime}$.5 and the B2 shocked region at $\alpha$(J2000)~=~20$^{\mbox{h}}$39$^{\mbox{m}}$13$^{\mbox{s}}$, $\delta$(J2000)~=~68$^{\circ}$00$^{\prime}$37$^{\prime\prime}$ (see Figure \ref{methanolcsobeam}) and spectra were adjusted to a $v_{LSR}$~=~1.75~km~s$^{-1}$.   A small subset of scans toward B1 were obtained at a second position offset by~-4\arcsec~in declination. The CSO 230/460 GHz sidecab double-sideband (DSB) heterodyne receiver was used in fair weather ($\tau_{220} =$~0.1~--~0.25) which resulted in typical system temperatures of 300~--~500~K.  The backend was a fast Fourier-transform spectrometer which provided 1 GHz of DSB data at 122 kHz resolution ($\sim$0.2~km~s$^{-1}$ at 230~GHz).  The total frequency coverage was 188 -- 193 GHz, 200 -- 205 GHz, 237 -- 243 GHz, and 249 -- 255 GHz toward B1, and 237 -- 243 GHz and 249 -- 255 GHz toward B2.  The targeted transitions of NH$_2$OH are given in Table \ref{NH2OHparams}.

A chopping secondary mirror with a throw of 4\arcmin~ was used for ON-OFF calibration.  Pointing was performed every $\sim$2 hours and typically converged to within 1~--~2\arcsec.  The raw data were calibrated using the standard chopper wheel calibration method resulting in intensities on the atmosphere-corrected $T_A^*$ temperature scale.  These were then corrected to the main beam temperature scale, $T_{mb}$, where $T_{mb} = T_A^*/\eta_{mb}$.  For these observations, the main beam efficiency was $\eta_b = 0.70$.  The spectra were collected in DSB mode at a variety of IF settings to allow for a robust deconvolution.  The CLASS package from the GILDAS suite of programs\footnote{Institut de Radioastronomie Millim\'{e}trique, Grenoble, France - http://www.iram.fr/IRAMFR/GILDAS} was used for the data reduction.  Spurious signals were removed from the spectra, which were then baseline subtracted using a polynomial fit.  The standard CLASS deconvolution routine was used to generate single-sideband data.  The spectra were then Hanning smoothed to a resolution of $\sim$1.4~km~s$^{-1}$.  The average full-width half-maximum linewidth in B1 was $\sim$7.9~km~s$^{-1}$, and in B2 was $\sim$5.2~km~s$^{-1}$.

\begin{deluxetable}{l c c c c c c}
\tablecolumns{7}
\tabletypesize{\footnotesize}
\tablecaption{Targeted NH$_2$OH transitions, spectroscopic parameters, assumed collisional coefficients, and critical densities.}
\tablewidth{0pt}
\tablehead{
\colhead{Transition}		&	\colhead{Frequency\tablenotemark{*}}		&	\colhead{$E_U$}	&	\colhead{S$_{ij}\mu^2$}	&	\colhead{$\gamma$\tablenotemark{\dagger}}			&	\colhead{log($A$)}		&	\colhead{$n_{cr}$\tablenotemark{\ddagger}} \\
\colhead{}				&	\colhead{(MHz)}		&	\colhead{(K)}		&	\colhead{(Debye$^2$)}	&	\colhead{(cm$^3$ s$^{-1}$)}	&	\colhead{(log(s$^{-1}$))}	&	\colhead{(cm$^{-3}$)}
}
\startdata
2$_{1,2}$ -- 1$_{1,1}$	&	100683.58(20)			&	15.204			&	0.520				&	$4.1 \times 10^{-11}$		&	-5.9079				&	$3 \times 10^{4}$\\
2$_{0,2}$ -- 1$_{0,1}$	&	100748.23(20)			&	7.2527			&	0.694				&	$5.5 \times 10^{-11}$		&	-5.7821				&	$3 \times 10^{4}$\\
2$_{1,1}$ -- 1$_{1,0}$	&	100807.62(20)			&	15.213			&	0.520				&	$4.1 \times 10^{-11}$		&	-5.9062				&	$3 \times 10^{4}$\\
\vspace{-0.5em} & \\
5$_{1,5}$ -- 4$_{1,4}$	&	251677.3666(78)		&	44.194			&	1.665				&	$4.0 \times 10^{-12}$		&	-4.5515				&	$7 \times 10^{6}$\\
5$_{4,1}$ -- 4$_{4,0}$	&	251734.8061(82)		&	163.51			&	0.625				&	$2.4 \times 10^{-11}$		&	-4.9771				&	$4 \times 10^{5}$\\
5$_{4,2}$ -- 4$_{4,1}$	&	251734.8061(82)		&	163.51			&	0.625				&	$2.4 \times 10^{-11}$		&	-4.9771				&	$4 \times 10^{5}$\\
5$_{3,3}$ -- 4$_{3,2}$	&	251780.3167(70)		&	107.85			&	1.110				&	$1.6 \times 10^{-11}$		&	-4.7270				&	$1 \times 10^{6}$\\
5$_{3,2}$ -- 4$_{3,1}$	&	251780.3178(70)		&	107.85			&	1.110				&	$1.6 \times 10^{-11}$		&	-4.7270				&	$1 \times 10^{6}$\\
5$_{2,4}$ -- 4$_{2,3}$	&	251811.9913(70)		&	68.081			&	1.457				&	$2.0 \times 10^{-13}$		&	-4.6088				&	$1 \times 10^{8}$\\
5$_{2,3}$ -- 4$_{2,2}$	&	251813.8609(70)		&	68.081			&	1.457				&	$2.0 \times 10^{-13}$		&	-4.6087				&	$1 \times 10^{8}$\\
5$_{0,5}$ -- 4$_{0,4}$	&	251838.4937(79)		&	36.261			&	1.735				&	$1.2 \times 10^{-10}$		&	-4.5329				&	$1 \times 10^{5}$\\
5$_{1,4}$ -- 4$_{1,3}$	&	251987.1350(78)		&	44.239			&	1.665				&	$4.0 \times 10^{-12}$		&	-4.5498				&	$1 \times 10^{6}$\\
\enddata
\tablenotetext{*}{Transitions and parameters accessible at www.splatalogue.net.  Original laboratory work reported by \citet{Tsunekawa1972} and \citet{Morino2000}. Catalogued at CDMS \citep{Muller2005}.}
\tablenotetext{\dagger}{Values were taken from corresponding transitions of $A$-CH$_3$OH obtained from the Leiden Atomic and Molecular Database \citep{Schoier2005} and used without further modification.  Original data reported by \citet{Rabli2010}.}
\tablenotetext{\ddagger}{Given as $n_{cr}$ = $A/\gamma$}
\label{NH2OHparams}
\end{deluxetable}

\section{CARMA Observations}

A total of 89.3 hours of observations were conducted with the CARMA 15-element array in C-configuration (2013 May), D-configuration (2012 October, November), and E-configuration (2012 August) at $\lambda = 3$ mm.  The phase center for these observations was $\alpha$(J2000) = 20$^{\mbox{h}}$39$^{\mbox{m}}$07$^{\mbox{s}}$.7, $\delta$(J2000) = 68$^{\circ}$01$^{\prime}$11$^{\prime\prime}$.5.  The CARMA correlator was used in its 62 MHz bandwidth, 3-bit mode providing 255 channels across the band for a resolution of 243~kHz or 0.7~km~s$^{-1}$.  The three targeted transitions of NH$_2$OH are given in Table \ref{NH2OHparams}.

MWC349 and Neptune were used as primary flux calibrators; the passband calibrators were 1635+381, 2232+117, 0102+584, 1743-038, 2015+372, and 3C84.  The gain calibrator was 1927+739.  Data reduction was completed using the Miriad package. Data were flagged for antennas that were offline or malfunctioning during observation or when the phases in the calibrators had deviations greater than thirty degrees from a smooth trend line. Image cleaning was also performed using Miriad. The robust factor tended towards natural weighting with cell size set to 0.4\arcsec~and image size of 1024 pixels. No channel averaging was applied for the maps shown; however, channel averaging to 1.4 km s$^{-1}$ was used on windows targeting hydroxylamine transitions. Clean regions were drawn with the polygon tool around clear emission and cleaned to an average noise of 0.7~mJy beam$^{-1}$. The restoring beam was typically $\sim$3\arcsec.4 $\times$ 3\arcsec.2.

The full set of observations toward L1157 CARMA included a host of other molecular species, and will be presented in a follow-up paper by Dollhopf et al. (2015).

\section{Non-LTE Modeling of CH$_3$OH}

An accurate determination of NH$_2$OH column density upper limits in the B1 and B2 shocks requires the kinetic temperature and density in these regions. To obtain these, we fit the observed CH$_3$OH emission in the same spectral window (Table \ref{CH3OHparams}) to a non-LTE model of the outflow and shock, based on a physical model derived from our our CARMA observations, using the RADEX code \citep{vandertak2007}.

\begin{deluxetable}{l c c c c c c}
\tablecolumns{7}
\tabletypesize{\footnotesize}
\tablecaption{Observed CH$_3$OH transitions, spectroscopic parameters, collisional coefficients, and critical densities.}
\tablewidth{0pt}
\tablehead{
\colhead{Transition\tablenotemark{\triangle}}		&	\colhead{Frequency\tablenotemark{*}}		&	\colhead{$E_U$}	&	\colhead{S$_{ij}\mu^2$}	&	\colhead{$\gamma$\tablenotemark{\dagger}}			&	\colhead{log($A$)}		&	\colhead{$n_{cr}$\tablenotemark{\ddagger}} \\
\colhead{}				&	\colhead{(MHz)}		&	\colhead{(K)}		&	\colhead{(Debye$^2$)}	&	\colhead{(cm$^3$ s$^{-1}$)}	&	\colhead{(log(s$^{-1}$))}	&	\colhead{(cm$^{-3}$)}
}
\startdata
\multicolumn{7}{c}{$A$-CH$_3$OH} \\
\vspace{-0.5em} & \\
2$_{0,2}$ -- 1$_{0,1}$ + +		&	96741.375(5)	&	6.9650	&	1.617	&	$5.5 \times 10^{-11}$	&	-5.4675	&	$6 \times 10^{4}$	\\
\vspace{-0.5em} & \\
5$_{1,5}$ -- 4$_{1,4}$ + +		&	239746.253	&	49.059	&	3.885	&	$4.3 \times 10^{-12}$	&	-4.2468	&	$1 \times 10^{7}$	\\
5$_{0,5}$ -- 4$_{0,4}$ + +		&	241791.431	&	34.817	&	4.043	&	$1.2 \times 10^{-10}$	&	-4.2184	&	$5 \times 10^{5}$	\\      
5$_{3,3}$ -- 4$_{3,2}$ + +		&	241832.91(20)	&	84.618	&	2.578	&	$1.6 \times 10^{-11}$	&	-4.4137	&	$2 \times 10^{6}$	\\
5$_{3,2}$ -- 4$_{3,1}$ - -		&	241832.91(20)	&	84.618	&	2.578	&	$4.6 \times 10^{-11}$	&	-4.4137	&	$8 \times 10^{5}$	\\
5$_{2,4}$ -- 4$_{2,3}$ - -		&	241842.324	&	72.530	&	3.415	&	$1.3 \times 10^{-13}$	&	-4.2915	&	$4 \times 10^{8}$	\\
5$_{2,3}$ -- 4$_{2,2}$ + +		&	241887.704	&	72.533	&	3.415	&	$1.8 \times 10^{-13}$	&	-4.2912	&	$3 \times 10^{8}$	\\
5$_{1,4}$ -- 4$_{1,3}$ - -		&	243915.826	&	49.661	&	3.885	&	$3.8 \times 10^{-13}$	&	-4.2243	&	$2 \times 10^{8}$	\\
11$_{0,11}$ -- 10$_{1,10}$ + +	&	250506.980	&	153.10	&	10.63	&	$9.0 \times 10^{-13}$	&	-4.0728	&	$9 \times 10^{7}$	\\
8$_{3,5}$ -- 8$_{2,6}$ - +		&	251517.262	&	133.36	&	7.308	&	$8.1 \times 10^{-13}$	&	-4.0990	&	$1 \times 10^{8}$	\\
7$_{3,4}$ -- 7$_{2,5}$ - +		&	251641.667	&	114.79	&	6.279	&	$1.4 \times 10^{-12}$	&	-4.1099	&	$6 \times 10^{7}$	\\
6$_{3,3}$ -- 6$_{2,4}$ - +		&	251738.520	&	98.546	&	5.224	&	$6.7 \times 10^{-12}$	&	-4.1272	&	$1 \times 10^{7}$	\\
5$_{3,2}$ -- 5$_{2,3}$ - +		&	251811.882	&	84.618	&	4.126	&	$1.8 \times 10^{-11}$	&	-4.1567	&	$4 \times 10^{6}$	\\
4$_{3,1}$ -- 4$_{2,2}$ - +		&	251866.579	&	73.012	&	2.954	&	$2.2 \times 10^{-11}$	&	-4.2144	&	$3 \times 10^{6}$	\\
5$_{3,3}$ -- 5$_{2,4}$ + -		&	251890.901	&	84.619	&	4.125	&	$8.1 \times 10^{-12}$	&	-4.1564	&	$9 \times 10^{6}$	\\
6$_{3,4}$ -- 6$_{2,5}$ + -		&	251895.728	&	98.545	&	5.220	&	$2.6 \times 10^{-12}$	&	-4.1266	&	$3 \times 10^{7}$	\\
4$_{3,2}$ -- 4$_{2,3}$ + -		&	251900.495	&	73.012	&	2.953	&	$4.5 \times 10^{-11}$	&	-4.2143	&	$1 \times 10^{6}$	\\
3$_{3,0}$ -- 3$_{2,1}$ - +		&	251905.812	&	63.727	&	1.641	&	$3.1 \times 10^{-11}$	&	-4.3603	&	$1 \times 10^{6}$	\\
3$_{3,1}$ -- 3$_{2,2}$ + -		&	251917.042	&	63.727	&	1.641	&	$1.3 \times 10^{-12}$	&	-4.3603	&	$3 \times 10^{7}$	\\
7$_{3,5}$ -- 7$_{2,6}$ + -		&	251923.631	&	114.79	&	6.272	&	$8.5 \times 10^{-11}$	&	-4.1089	&	$9 \times 10^{5}$	\\
8$_{3,6}$ -- 8$_{2,7}$ + -		&	251984.702	&	133.36	&	7.293	&	$2.1 \times 10^{-11}$	&	-4.0975	&	$4 \times 10^{6}$	\\
\vspace{-0.5em} &\\
\multicolumn{7}{c}{$E$-CH$_3$OH} \\
\vspace{-0.5em} &\\
2$_{-1,2}$ -- 1$_{-1,2}$ 		&	96739.362(5)	&	12.541	&	1.213	&	$1.4 \times 10^{-13}$	&	-5.5922	&	$2 \times 10^{7}$	\\
2$_{0,2}$ -- 1$_{0,2}$ 		&	96744.550(5)	&	20.090	&	1.617	&	$9.5 \times 10^{-12}$	&	-5.4675	&	$4 \times 10^{5}$	\\
2$_{1,1}$ -- 1$_{1,0}$ 		&	96755.511(5)	&	28.011	&	1.244	&	$3.7 \times 10^{-11}$	&	-5.5811	&	$7 \times 10^{4}$	\\
\vspace{-0.5em} & \\
5$_{0,5}$ -- 4$_{0,4}$ 		&	241700.219	&	47.934	&	4.040	&	$3.0 \times 10^{-11}$	&	-4.2192	&	$2 \times 10^{6}$	\\
5$_{-1,5}$ -- 4$_{-1,4}$ 		&	241767.224	&	40.391	&	3.882	&	$7.9 \times 10^{-13}$	&	-4.2362	&	$7 \times 10^{7}$	\\
5$_{3,2}$ -- 4$_{3,1}$ 		&	241843.646	&	82.531	&	2.587	&	$1.5 \times 10^{-12}$	&	-4.4120	&	$3 \times 10^{7}$	\\    
5$_{1,4}$ -- 4$_{1,3}$ 		&	241879.073	&	55.871	&	3.980	&	$3.1 \times 10^{-12}$	&	-4.2248	&	$2 \times 10^{7}$	\\
5$_{-2,4}$ -- 4$_{-2,3}$ 		&	241904.152	&	60.725	&	3.399	&	$1.1 \times 10^{-13}$	&	-4.2932	&	$5 \times 10^{8}$	\\
5$_{2,3}$ -- 4$_{2,2}$ 		&	241904.645	&	57.069	&	3.356	&	$2.5 \times 10^{-12}$	&	-4.2987	&	$2 \times 10^{7}$	\\      
2$_{0,2}$ -- 1$_{-1,1}$ 		&	254015.340	&	20.090	&	0.499	&	$4.7 \times 10^{-13}$	&	-4.7208	&	$4 \times 10^{7}$	\\      
\enddata
\tablecomments{Only transitions with modeled peak intensity above 5 mK are listed here.  A complete modeled spectrum is available as Supplementary Information.}
\tablenotetext{\triangle}{``+" and ``-" refer to $A+$ and $A-$ parity states, respectively. For $E$-CH$_3$OH, a negative value of $K_a$ is used to differentiate between (+)$E_1$ and (-)$E_2$ states, which belong to the same $E$ symmetry species.}
\tablenotetext{*}{Transitions and parameters accessible at www.splatalogue.net.  Original laboratory work reported by \citet{Xu1997} and references therein and by \citet{Muller2004}. Catalogued at CDMS \citep{Muller2005}.  Except where noted, uncertainties are 50 kHz.}
\tablenotetext{\dagger}{Values obtained from the Leiden Atomic and Molecular Database \citep{Schoier2005}.  Original data reported by \citet{Rabli2010}.}
\tablenotetext{\ddagger}{Given as $n_{cr}$ = $A/\gamma$}
\label{CH3OHparams}
\end{deluxetable}

\begin{figure}
\centering
\plotone{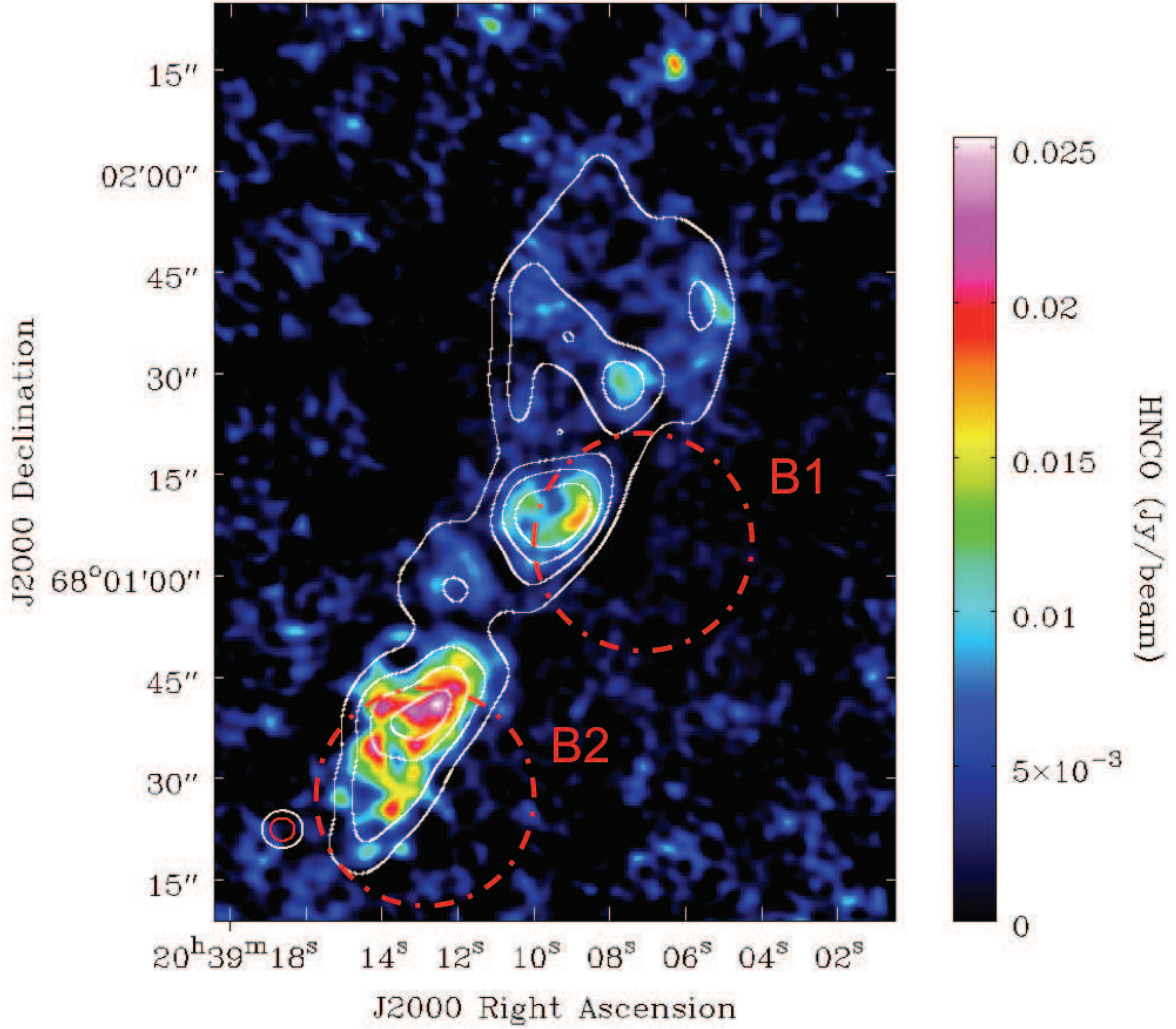}
\caption{CARMA observations of CH$_3$OH and HNCO toward L1157.  Contours are an integrated Moment 0 map of the $2_{-1,2} - 1_{-1,1}$,  $2_{0,2} - 1_{0,1}$++, and $2_{0,2} - 1_{0,1}$ transitions ($E_u =$ 12.5 K, 7 K, and 20.1 K, respectively) of CH$_3$OH at contour levels of 0.16, 0.32, 0.49, 0.66, and 0.82 Jy/beam.  These are overlaid on an integrated Moment 0 map of the $4_{0,4} - 3_{0,3}$ transition of HNCO spanning 0 $\rightarrow$ 0.025 Jy/beam in a single color cycle.  The HNCO transitions trace the warmer, compact shocks, while the CH$_3$OH transitions reveal the overall structure of the colder, diffuse outflow. The location of the two targeted observations of B1 and B2 with the CSO are shown as dashed white lines, approximately equal to the CSO beam size ($\sim$30\arcsec) at the observed frequencies.}
\label{methanolcsobeam}
\end{figure}

First, we determined the approximate spatial extent and location of the shocked gas from the CH$_3$OH emission observed with CARMA.  Figure \ref{methanolcsobeam} shows the CSO beam at the targeted positions overlaid on our CARMA observations of CH$_3$OH and HNCO\footnote{Details and analysis of the HNCO observations will be presented in Paper II by Dollhopf et al. (2015)} at $\lambda = 3$ mm.  The pointing position for B1 was chosen from the literature prior to our CARMA observations, and is roughly centered on the shock front, rather than the peak of the shock emission.  The pointing position for B2 was chosen after we had acquired the CARMA observations, and is centered on the peak of the shocked HNCO emission.  In both cases, the shock is nearly completely contained within the beam, and we estimate a reasonable equivalent size of $\sim$12\arcsec~ for the shock, and $\sim$20\arcsec~ for the second component, within our $\sim$30\arcsec~ CSO beam.  In B1, the shock is not centered in the beam, and thus any emission will suffer more significant fall-off effects than in B2.

We made the initial assumption that the two components in our fit represented the shocked gas and the outflow.  Under these assumptions, and based on our CARMA observation as well as prior studies \citep{Bachiller1995}, we constrained the fit of the CH$_3$OH emission with the following assumptions:

\begin{enumerate}

\item The kinetic temperature of the shock was no less than that of the outflow.
\item The density of the shock was no less than that of the outflow.
\item The column density of CH$_3$OH was no less in the shock than in the outflow.

\end{enumerate}

Assumptions (1) and (2) are well-substantiated in the literature, both toward L1157 \citep{Bachiller1995} and in shocked environments in general \citep{vanDishoeck1998}, and assumption (3) agrees with the previous findings of \citet{Bachiller1995}.  Using these constraints, and the two-source component model described earlier, we performed a reduced-$\chi^2$ analysis of a grid of models for the CSO observations, simultaneously fitting $A$- and $E$-CH$_3$OH following the methods described in \citet{Crockett2014}.  A thorough discussion of the uncertainties in the fitting method is provided in Appendix \ref{appA}.

Based on these uncertainties, and the assumptions described above, we find the parameters in Table \ref{methanolradex} represent a best-fit to the data, with reduced-$\chi^2$ values of 2.50 and 1.77 for B1 and B2, respectively.  These values are likely due to the simplicity of our model relative to the complexity of the source.  Literature sources (see, e.g., \citet{Lefloch2012}) suggest a structure with three or more components is more realistic, but such additional components would not be well-constrained in our case due to the limited number of observed CH$_3$OH transitions.  Additionally, there are likely quite large temperature and density gradients, especially in the newer, more compact B1 shock, which will contribute additional error to the fit which is not accounted for in our analysis.  Finally, CH$_3$OH excitation is sensitive to the far-IR radiation field present, which we have assumed as standard galactic background, any deviation from which would further impact the accuracy of the model and increased the ultimate $\chi^2$.

\begin{figure}
\centering
\plotone{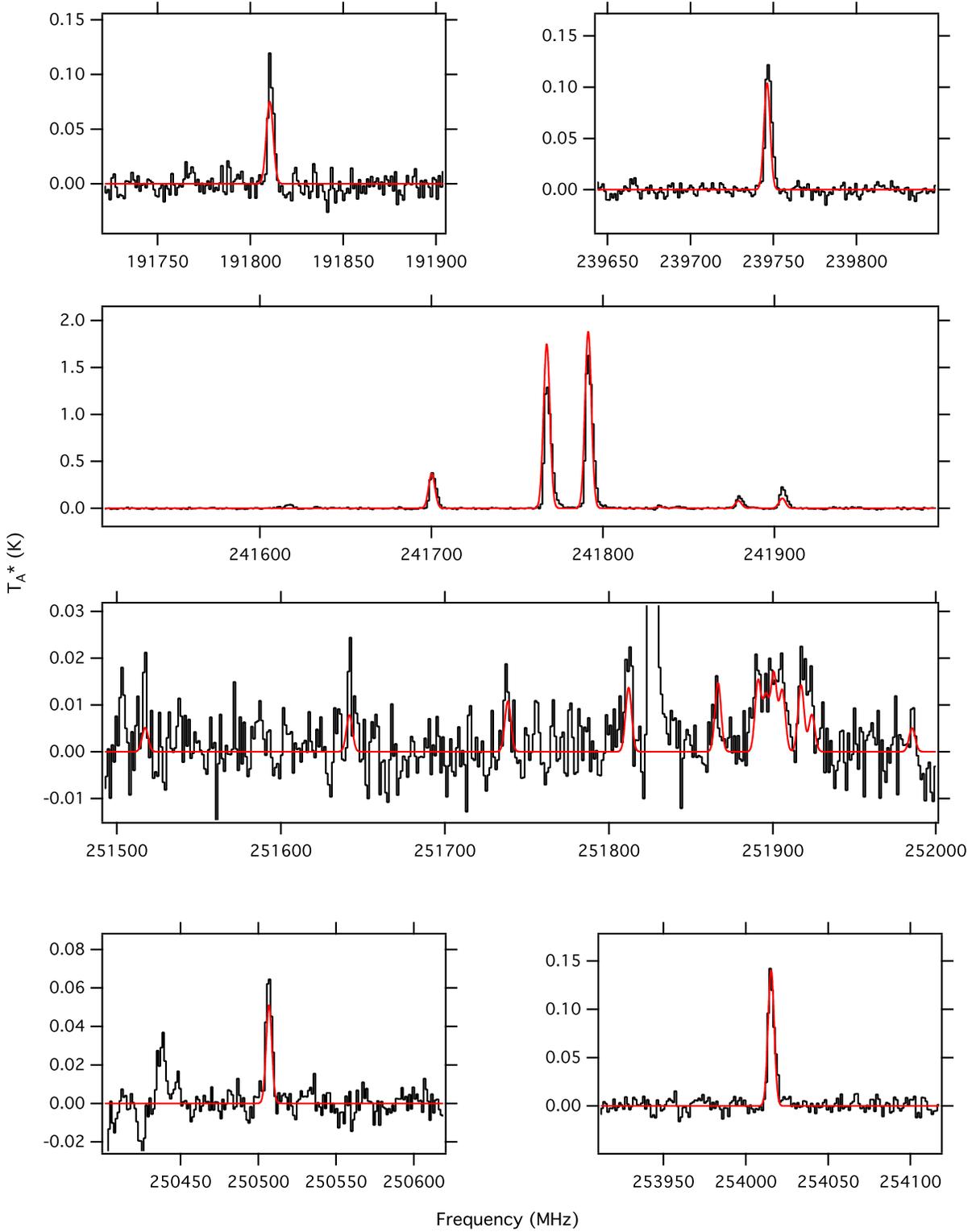}
\caption{Non-LTE simulation of CH$_3$OH spectrum from RADEX fit in red, overlaid on CSO observations of the B1 shock in black.  The $4_{1,4} - 3_{1,3}$$^{++}$ transition of $A$-CH$_3$OH at 191810.5 MHz was not included as part of the RADEX fit.  It is, however, reasonably well-reproduced by the fit of the higher-frequency lines, adding additional evidence for the robustness of the fit.}
\label{methanolresults}
\end{figure}

\begin{figure}
\centering
\plotone{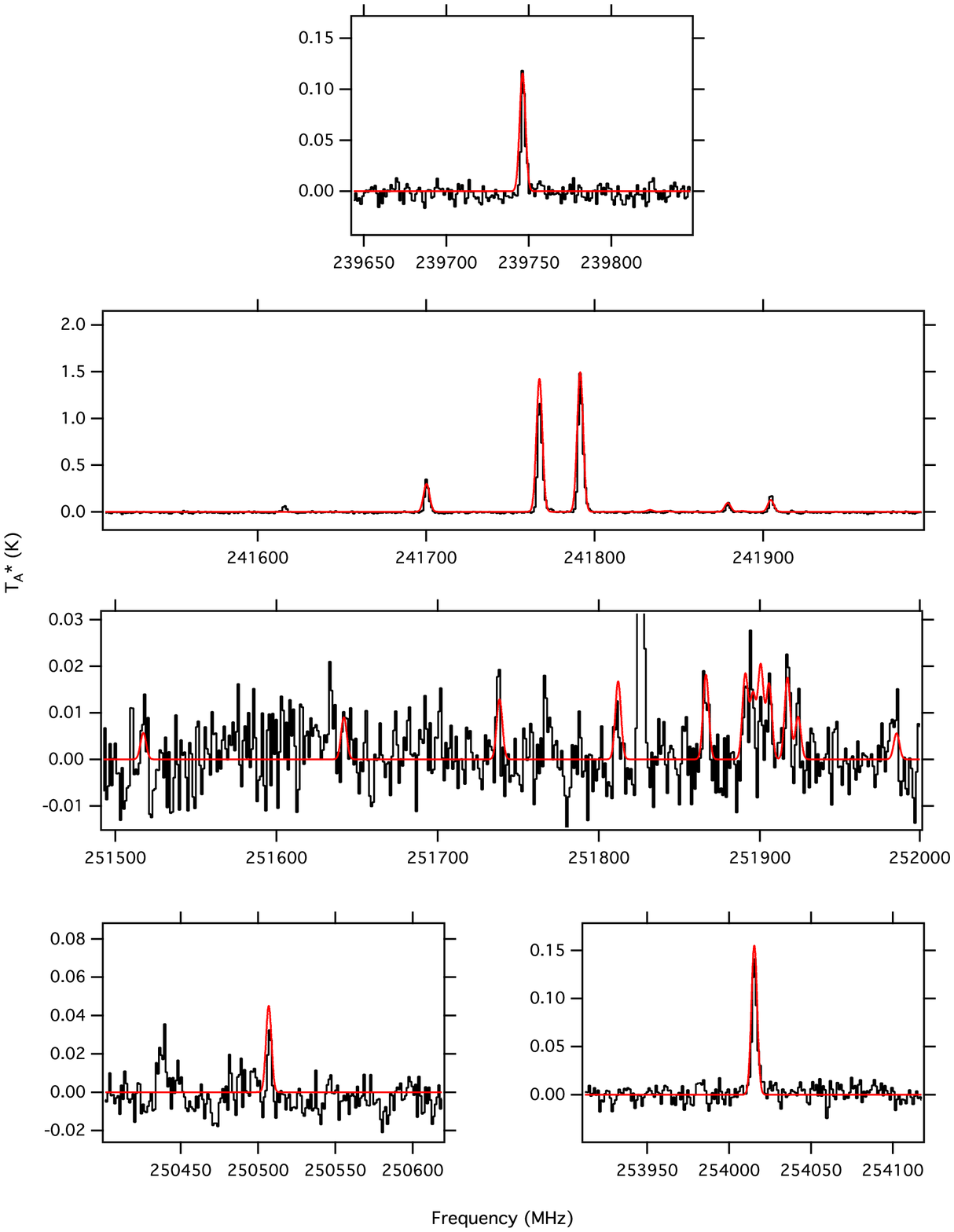}
\caption{Non-LTE simulation of CH$_3$OH spectrum from RADEX fit in red, overlaid on CSO observations of the B2 shock in black.}
\label{methanolresultsb2}
\end{figure}

Simulated non-LTE spectra of CH$_3$OH from these results toward B1 and B2 are provided in Figures \ref{methanolresults} and \ref{methanolresultsb2}, respectively, overlaid on observations and corrected for beam efficiency and a 12\arcsec~source size.  The CH$_3$OH column densities and H$_2$ densities found here are in relatively good agreement with previous observations of the source \citep{Bachiller1995,Bachiller1997,Sugimura2010}. The derived values of $T_{kin}$ for the warmer component, which we ascribe to the shocks, agrees well with previous measurements in B2 \citep{Lefloch2012}, while the upper range of our derived value in B1 falls at the lower edge of previous measurements.

The kinetic temperature of the second component ($T_{kin} = 10$ K) is significantly lower than previous measurements of both the shock and the outflow ($T_{kin} \sim 60 - 120$ K), indicating that it is likely not probing the outflow as we had originally presumed.  Instead, toward B1, this component is more likely related to the $g_3$ component described by \citet{Lefloch2012} as remnants from the gas in which the earlier B2 shock was formed, also offering a possible explanation for the component's presence toward B2.  The $g_3$ component has been reported to be both cold ($T_{kin} = 23$ K) and extended \citet{Mendoza2014,Lefloch2012}, in qualitatively good agreement with our fit.

\begin{deluxetable}{c c c c c c }
\tablecolumns{3}
\tabletypesize{\footnotesize}
\tablecaption{Best-fit parameters for two components of CH$_3$OH toward L1157-B1 and B2 determined from non-LTE RADEX calculations.}
\tablewidth{0pt}
\tablehead{
\colhead{}						&	\multicolumn{2}{c}{L1157-B1}	& \hspace{0.25in} & \multicolumn{2}{c}{L1157-B2} \\
\cline{2-3} \cline{5-6}
\colhead{Parameter}				&	\colhead{Component 1}	&	\colhead{Component 2} &	& \colhead{Component 1}	&	\colhead{Component 2}
}
\startdata
$n_{H_2}$ (cm$^{-3}$)			&	$3(1) \times 10^5$			&	$3(1) \times 10^5$			&&	$6(2) \times 10^5$			&	$6(2) \times 10^5$			\\
$T_{kin}$ (K)					&	60(20)					&	10(3)						&&	50(15)					&	10(3)						\\
$N_{CH_3OH}$ (cm$^{-2}$)		&	$3(1) \times 10^{15}$		&	$5(2) \times 10^{14}$		&&	$3(1) \times 10^{15}$		&	$6(2) \times 10^{14}$		\\
\enddata
\tablecomments{Errors given are from Equation \ref{uncertainty} and calculated to be $\sim$32\%.}
\label{methanolradex}
\end{deluxetable}

\section{Results}
\label{results}

Despite deep searches, we find no conclusive evidence for NH$_2$OH emission at $\lambda = 3$~mm or $\lambda = 1.3$~mm in either the CARMA or the CSO data, respectively.  To calculate appropriate upper limits for NH$_2$OH in the both datasets, we assume that NH$_2$OH, if present in the gas phase, originates in the warm, shocked gas traced by the warm, dense CH$_3$OH component.  The rationale for this assumption is discussed in \S\ref{discussion}.

\subsection{NH$_2$OH in CSO Data}

Near the temperatures derived from our RADEX fit in the warm component toward B1 and B2, the strongest NH$_2$OH transition is the 5$_{0,5}$ -- 4$_{0,4}$ at 251838 MHz.  As noted in Table \ref{NH2OHparams}, the critical density ($n_{cr}$) for this transition is $10^5$~cm$^{-3}$. This value is based on the assumption that the collisional coefficients for NH$_2$OH are similar to those of $A$-CH$_3$OH.  While not exact, the similar mass, dipole moments, molecular size, and energy level structures make this a reasonable approximation within the context of the following discussion.

Given the densities derived from our RADEX fit (3 -- 6 $\times 10^5$ cm$^{-3}$), we therefore assume that, at least for this transition,  LTE is a reasonable approximation for determining NH$_2$OH upper limits and thus $T_{ex} = T_{kin}$.  Under these conditions, we derive 1$\sigma$ upper limit column densities for NH$_2$OH of $\leq1.4 \times 10^{14}$~cm$^{-2}$ and $\leq1.0 \times 10^{14}$~cm$^{-2}$ in B1 and B2, respectively.

\citet{Lefloch2012} derive a CO column density of $0.9 \times 10^{17}$~cm$^{-2}$ in the $g_2$ component of the B1 shock, arising from the shocked gas and covering the entire region, assuming a 20\arcsec~source size.  Assuming the CO is homogeneously distributed over the region, this gives a source-averaged CO column of $3.2 \times 10^{16}$~cm$^{-2}$ for the 12\arcsec~source size used here.  Taking $N_{H_2} = 3 \times 10^4 (N_{CO})$ (c.f. \citet{Bolatto2013} and \citet{Lefloch2012}), this results in an H$_2$ column density~of~$\sim$$10^{21}$~cm$^{-2}$.   A similar argument using the CO column density of $5.2 \times 10^{16}$~cm$^{-2}$ for B2 found by \cite{Bachiller1997} also results in an H$_2$ column density~of~$\sim$$10^{21}$~cm$^{-2}$.  This gives upper limits to the fractional abundances of NH$_2$OH of $N_{NH_2OH}/N_{H_2} = \leq1.4 \times 10^{-7}$ and $\leq1.0 \times 10^{-7}$ for B1 and B2, respectively.

\subsection{NH$_2$OH in CARMA Data}

\begin{figure}
\centering
\plotone{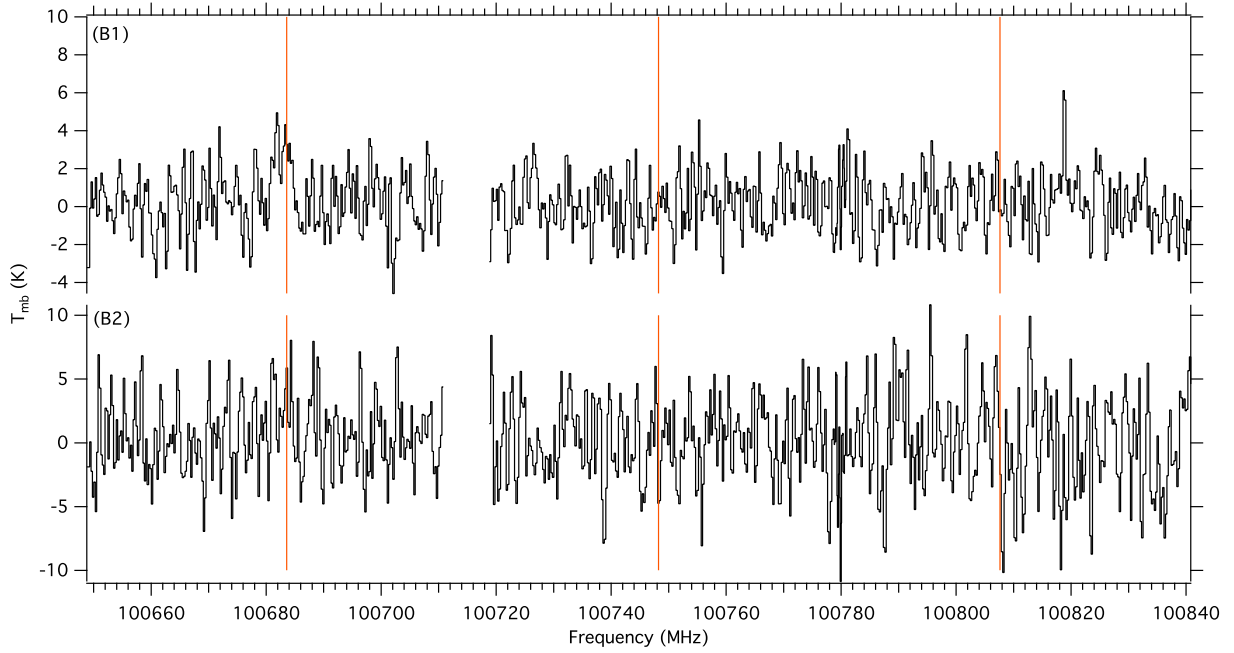}
\caption{Spectra toward L1157-B1 (bottom) and B2 (top) extracted from 12\arcsec~beams centered on the peaks of the shocked CH$_3$OH emission and smoothed to a velocity resolution of $\sim$1.4 km s$^{-1}$.  Red lines indicate the frequencies of NH$_2$OH transitions in this frequency region.  A U-line is present toward the B1 shock at the frequency of the $2_{1,2} - 1_{1,1}$ NH$_2$OH transition.}
\label{nh2ohcarma}
\end{figure}

To calculate appropriate upper limits for NH$_2$OH in the CARMA data, we again assume that NH$_2$OH, if present in the gas phase, originates in the warm, shocked gas traced by the warm, dense CH$_3$OH component.  The spectra extracted from a 12\arcsec~beam centered on these locations are shown in Figure \ref{nh2ohcarma}.  While there is an unidentified feature at the frequency of the $2_{1,2} - 1_{1,1}$ NH$_2$OH transition, if this were truly NH$_2$OH emission, the other transitions in the observed window would be of equal or greater intensity.

We measure the RMS to be 1.6 mK and 2.9 mK in the B1 and B2 spectra.  At the derived temperatures, the $2_{0,2} - 1_{0,1}$ transitions at 100748 MHz is predicted to be strongest.  As both B1 and B2 have densities higher than $n_{cr}$ for this transition (see Table \ref{NH2OHparams}), we again derive an upper limit to the column density assuming LTE conditions.

Following the same procedure as for the CSO data, we find a 1$\sigma$ upper limit of NH$_2$OH of $\leq 1.4 \times 10^{13}$ cm$^{-2}$ for B1 and $\leq 1.5 \times 10^{13}$ cm$^{-2}$ for B2.  This gives upper limits to the fractional abundances of NH$_2$OH of $N_{NH_2OH}/N_{H_2} = \leq1.4 \times 10^{-8}$ and $\leq1.5 \times 10^{-8}$ for B1 and B2, respectively.  All results are summarized in Table \ref{NH2OHcdresults}.

\begin{deluxetable}{c c c c c c}
\tablecolumns{5}
\tabletypesize{\footnotesize}
\tablecaption{Derived upper limits to NH$_2$OH column density and relative abundance in L1157-B1 and L1157-B2 from the CSO and CARMA data.}
\tablewidth{0pt}
\tablehead{
				&	\multicolumn{2}{c}{CSO $\lambda = 1$ mm}	& \hspace{0.15in}	&	\multicolumn{2}{c}{CARMA $\lambda = 3$ mm} \\
\cline{2-3} \cline{5-6}				
				&	\colhead{B1}	&	\colhead{B2}			&	&	\colhead{B1}	&	\colhead{B2}			
}
\startdata
$N_{NH_2OH}$ (cm$^{-2}$)	&	$\leq1.4 \times 10^{14}$	& $\leq1.0 \times 10^{14}$		&	&	$\leq 1.4 \times 10^{13}$	&	$\leq 1.5 \times 10^{13}$ \\
$N_{NH_2OH}/N_{H_2}$		&	$\leq1.4 \times 10^{-7}$	& $\leq1.0 \times 10^{-7}$		&	&	$\leq1.4 \times 10^{-8}$	&	$\leq1.5 \times 10^{-8}$	\\
\enddata
\label{NH2OHcdresults}
\end{deluxetable}

\section{Discussion and Conclusions}
\label{discussion}

In \S\ref{results}, we assumed that NH$_2$OH, if present, would arise predominantly the warm, shocked regions B1 or B2, rather than the molecular outflow.  \citet{Zheng2010} show that NH$_2$OH is thermally liberated from their laboratory samples between 160 -- 180 K, whereas for realistic interstellar ices and conditions, temperatures above $\sim$110 K are likely sufficient \citep{Collings2004}.  Both our non-LTE fits, and the literature values discussed earlier have shown that temperatures in the targeted regions are below these thresholds, and thus a thermal mechanism for the desorption of NH$_2$OH from grain surfaces in L1157 is unlikely.  Thus, non-thermal desorption in the shocks should be the dominant mechanism for NH$_2$OH liberation, and any NH$_2$OH emission should trace these shocked regions.

The upper limit column densities established with the CARMA data are equivalent to those established by \citet{Pulliam2012} of $\leq (0.9 - 8) \times 10^{13}$ cm$^{-2}$.  The limits established by \citet{Pulliam2012}, however, are beam-averaged column densities, whereas the CARMA measurements presented here provide a more well-constrained source size, and thus a more robust upper limit.  Nevertheless, the upper limits to relative abundance found here are significantly higher than those determined by \citet{Pulliam2012} of $\sim$10$^{-10}$ due to the low H$_2$ column in the region.  These low columns are common in other chemically-complex shocked regions as well, such as a host of Galactic Center clouds studied by \citet{Torres2006} ($N_{H_2} = (2 - 68) \times 10^{21}$ cm$^{-2}$), and other shocked outflows like BHR 71 ($N_{H_2} = (3 - 11) \times 10^{21}$ cm$^{-2}$; \citet{Garay1998}).

Yet, these regions likely represent the best-case scenario for a gas-phase detection of NH$_2$OH.   Such a detection is critical to accurately constrain the application of  laboratory results, which show NH$_2$OH is a significant player in grain-surface nitrogen chemistry, to chemical models.  The deep search presented here stretched the capabilities of the CSO and CARMA to the limit; searches deep enough to be sensitive to the relative abundances predicted by laboratory work and modeling are simply time-prohibitive on these types of instruments.  The \citet{Garrod2013} model, without the addition of the H + HNO $\rightarrow$ HNOH pathway, predicts peak condensed-phase NH$_2$OH abundances of $\sim$$(5-10) \times 10^{-9}$, which, if liberated \textit{en masse} in a shock, is a detectable population, but likely only with the sensitivity and spatial resolution of ALMA.  The upper limits established here, however, are several orders of magnitude lower than both the condensed-phase and gas-phase populations predicted by the augmented model which includes this hydrogenation pathway.  Thus, our results demonstrate that further efforts are needed both in modeling and in the laboratory, to identify and fully-constrain both the formation and the destruction pathways for NH$_2$OH.

\acknowledgments

BAM gratefully acknowledges funding by an NSF Graduate Research Fellowship during initial portions of this work.  The authors thank the anonymous referee for comments which increased the quality of this work.  We thank S. Radford for support of the CSO observations.  The National Radio Astronomy Observatory is a facility of the National Science Foundation operated under cooperative agreement by Associated Universities, Inc.  Portions of this material is based upon work at the Caltech Submillimeter Observatory, which was operated by the California Institute of Technology under cooperative agreement with the National Science Foundation (AST-0838261).  Support for CARMA construction was derived from the Gordon and Betty Moore Foundation, the Kenneth T. and Eileen L. Norris Foundation, the James S. McDonnell Foundation, the Associates of the California Institute of Technology, the University of Chicago, the states of California, Illinois, and Maryland, and the National Science Foundation.

\appendix
\section{Error Analysis}
\label{appA}

The uncertainty in the observations is given as Equation \ref{uncertainty}, modified from Equation A1 in \citet{Crockett2014}.

\begin{equation}
\sigma_{tot} = \sqrt{\sigma_{rms}^2 + \sigma_{cal}^2 + \sigma_{pt}^2 + \sigma_{bf}^2}
\label{uncertainty}
\end{equation}

$\sigma_{rms}$ is the uncertainty due to the RMS noise level of the spectra.  This value is consistently $\sim$0.0058 K across the CH$_3$OH transitions in B1, and $\sim$0.0078 K in B2.

$\sigma_{cal}$  is the uncertainty in the absolute flux calibration of the observations, which is $\sim$30\%.

$\sigma_{pt}$ is the uncertainty in the pointing accuracy, which is taken to be 2\arcsec~ based on the average convergence of pointing observations throughout the observing period.

$\sigma_{bf}$ is the uncertainty in the beam-filling factor.  The size of the telescope beam varies by $\sim$4\arcsec~ across the band, is well-described, and is calculated independently at each transition frequency, thus we assume no contribution to $\sigma_{bf}$ from the telescope beam.  However, the sizes of the two source components are not a varied parameter, and are only modestly well-described by our CARMA observations.  Further, in the case of B1, the shock is positioned off-center of the beam, making the emission more sensitive to fall-off effects that are not accounted for explicitly in the calculations.  We therefore assume a factor of two uncertainty in source size to take both issues into consideration, which is reflected in $\sigma_{bf}$.  

These result in net uncertainties of $\sim$32\% in the absolute intensity of the observed transitions, with the largest contribution from the absolute flux calibration.  $\chi^2$ values outside of the extremes represented by these uncertainties more than double those of the best-fit parameters, so we take $\sim$32\% to be a reasonable estimate of the overall uncertainty in n$_{H_2}$, $N_{CH_3OH}$, and $T_{kin}$.

\end{document}